\documentclass[11pt]{article}

\usepackage{amsmath,amssymb,amsfonts,bm}
\usepackage{mathtools}
\usepackage{cite}
\usepackage{geometry}
\title{Hadamard--Tetrahedral Symmetric Spacetime and Clifford-Like
Factorizations of Dirac and Maxwell Operators}
\author{Farhad Aghili\footnote{Affiliate Professor, Concordia University. E-mail: farhad.aghili@concordia.ca; farhad.aghili@gmail.com.}}
\date{}

\begin{document}
\maketitle

\begin{abstract}
A symmetric representation of Minkowski spacetime is constructed by
combining a normalized fourth-order Hadamard transformation with an
imaginary temporal coordinate. The resulting complex coordinates $q^a$
distribute time and space equally among four components and transform the
Minkowski metric into a Euclidean-form complex bilinear metric. The four
Hadamard spatial sign vectors form a regular tetrahedron and appear
simultaneously in the transformed Dirac and Maxwell operators. The Dirac
matrices satisfy a Euclidean-form Clifford algebra and contain tetrahedral
combinations of Pauli matrices, whereas the Riemann--Silberstein Maxwell
matrices realize the corresponding spin-1 structure. A gauge-invariant QED
Lagrangian is formulated in the same coordinates. In Fourier space, the
Dirac and Maxwell systems are governed by the common invariant
$\kappa_a\kappa^a$, while the complete Maxwell operator admits an exact
rectangular Clifford-like factorization. The formulation is further extended
to the weak-field Einstein--Dirac--Maxwell system. In harmonic gauge, the
linearized gravitational field is governed by the same scalar wave operator
$\Box_q$, while gravitational corrections to the Dirac and Maxwell equations
enter through the perturbed tetrad, spin connection, and curvature. This
common principal-operator structure suggests shared spectral kernels and
scalar-wave preconditioning for coupled field solvers.
\end{abstract}

\section{Introduction}

Minkowski spacetime is conventionally represented by coordinates
$x^\mu=(ct,x,y,z)$ and metric
$\eta_{\mu\nu}=\operatorname{diag}(-1,1,1,1)$. Although the distinction
between temporal and spatial directions is physically fundamental, the
diagonal form of the metric is representation dependent. Complexification
provides an alternative means of carrying the Lorentzian signature through
the coordinates and basis rather than through an explicitly indefinite
metric. Complex spacetime representations have a long history in relativistic
physics, including spinor, quaternionic, twistor, and complex-orthogonal
formulations \cite{PenroseRindler,Hestenes1966}. Recent studies continue to
examine complex Minkowski metrics, spin matrices, Lorentz algebras, and
related symmetry structures \cite{Mendes2024,Yousaf2026}.

A closely related development is the representation of electromagnetic and
relativistic wave equations by complex first-order matrix operators. The
Riemann--Silberstein field combines the electric and magnetic fields into a
complex vector and allows Maxwell's equations to be expressed in a form
resembling relativistic quantum wave equations
\cite{Silberstein1907,Oppenheimer1931,BialynickiBirula1996,
BialynickiBirula2013}. Majorana--Oppenheimer and subsequent representations
have further clarified the relation between Maxwell theory, Lorentz
representations, and Dirac-like matrix equations
\cite{Bogush2009,Ivashkevich2022,Khan2024}. Common matrix and numerical
formulations of the Maxwell and Dirac systems have also been developed
\cite{Bocker2018,Bao2004,Lorin2011,VandenBroeck2023}, while operator
factorization remains relevant in recent relativistic electrodynamics
\cite{Pedergnana2025}.

The present work investigates a synthesis generated by the normalized
fourth-order Hadamard transformation. The transformation distributes the
conventional temporal and spatial coordinates among four complex symmetric
coordinates $q^a$ and generates four spatial sign vectors corresponding to
the vertices of a regular tetrahedron. The same tetrahedral vectors organize
the transformed spin-$1/2$ Dirac matrices, the spin-1 Maxwell matrices, and
their dual spectral variables. The complete Maxwell operator, including the
divergence equation, admits a rectangular Clifford-like factorization
parallel to the Clifford factorization of the Dirac operator. The framework
is also extended consistently to linearized gravity: the harmonic-gauge
Einstein equation, the complete Maxwell system, and the squared free Dirac
operator share the same scalar principal operator in the symmetric
coordinates. The gravitational backreaction on the matter fields is retained
to first order through the tetrad, spin connection, and curvature terms.

The transformation $q^a=U^a{}_{\mu}x^\mu$ is linear, constant, and
invertible, and the formulation is an alternative representation of the
conventional relativistic field equations. Its primary interest is the
geometric, algebraic, and computational structure exposed by the
transformation. Greek indices $\mu,\nu=0,\ldots,3$ denote conventional
Minkowski components, Latin indices $a,b=1,\ldots,4$ denote symmetric
Hadamard-spacetime components, and $i,j,k=1,2,3$ denote ordinary spatial
components. Repeated indices are summed unless stated otherwise.

\section{Symmetric Hadamard Spacetime}

\subsection{Space-like symmetric representation}

Define the normalized fourth-order Hadamard matrix and the complex spacetime
transformation by
\begin{equation}
H_4=\frac12
\begin{bmatrix}
1&1&1&1\\
1&-1&1&-1\\
1&1&-1&-1\\
1&-1&-1&1
\end{bmatrix},
\qquad
U^a{}_{\mu}=
\left[H_4\operatorname{diag}(i,1,1,1)\right]^a{}_{\mu}.
\label{eq:Udef}
\end{equation}
Since $H_4^TH_4=I_4$, the transformation satisfies $U^\dagger U=I_4$ and
\begin{equation}
U^a{}_{\mu}\eta^{\mu\nu}U^b{}_{\nu}=\delta^{ab}.
\label{eq:metric-transform}
\end{equation}

The space-like symmetric Hadamard coordinates, denoted below by
$q_{\rm S}^a\equiv q^a$, are
\begin{equation}
q^a=U^a{}_{\mu}x^\mu
=\frac12
\begin{bmatrix}
ict+x+y+z\\
ict-x+y-z\\
ict+x-y-z\\
ict-x-y+z
\end{bmatrix}.
\label{eq:qdef}
\end{equation}
The inverse is $x^\mu=(U^{-1})^\mu{}_a q^a=(U^\dagger)^\mu{}_a q^a$,
and
\begin{equation}
\delta_{ab}q^aq^b
=\eta_{\mu\nu}x^\mu x^\nu
=-c^2t^2+x^2+y^2+z^2.
\label{eq:interval}
\end{equation}
The Euclidean appearance of the left-hand side does not imply Euclidean
physical spacetime because the coordinates $q^a$ are complex.

\subsection{Physical reality condition}

The physical Minkowski subspace is specified by
\begin{equation}
(q^a)^*=(R_q)^a{}_bq^b,
\qquad
R_q=I_4-2e_0e_0^T,
\qquad
e_0=\frac12(1,1,1,1)^T.
\label{eq:reality}
\end{equation}
Hence physical $q$ space is a four-real-dimensional slice of $\mathbb C^4$.
Equation~\eqref{eq:reality} is essential: it removes the additional degrees
of freedom that would arise if the four complex coordinates were treated as
independent.

\subsection{Time-like symmetric representation}

A complementary time-like symmetric representation is
\begin{equation}
q_{\rm T}^{a}=\frac12
\begin{bmatrix}
t+i(x+y+z)/c\\
t+i(-x+y-z)/c\\
t+i(x-y-z)/c\\
t+i(-x-y+z)/c
\end{bmatrix},
\label{eq:qtime}
\end{equation}
for which
\begin{equation}
\delta_{ab}q_{\rm T}^{a}q_{\rm T}^{b}
=t^2-\frac{x^2+y^2+z^2}{c^2}.
\end{equation}
Thus the two representations differ in which sector carries the imaginary
factor: $q_{\rm S}^a$ produces the mostly-plus interval, whereas
$q_{\rm T}^a$ produces its mostly-minus, time-normalized counterpart.
The remainder of the paper uses the space-like representation in
Eq.~\eqref{eq:qdef}.

\section{Tetrahedral Geometry and Lorentz Transformations}

\subsection{Regular-tetrahedron basis}

Introduce the four spatial sign vectors
\begin{equation}
\bm s^{\,1}=(1,1,1)^T,\quad
\bm s^{\,2}=(-1,1,-1)^T,\quad
\bm s^{\,3}=(1,-1,-1)^T,\quad
\bm s^{\,4}=(-1,-1,1)^T.
\label{eq:s-vectors}
\end{equation}
They satisfy
\begin{equation}
\bm s^{\,a}\cdot\bm s^{\,b}=4\delta^{ab}-1.
\label{eq:tetra}
\end{equation}
Thus $\bm s^{\,a}/\sqrt3$ point toward the vertices of a regular
tetrahedron. The Hadamard basis vectors are $e_0^a=1/2$ and
$e_i^a=s_i^{\,a}/2$, so that $q^a=ict\,e_0^a+x^ie_i^a$.

\subsection{Lorentz transformations on the physical slice}

For $x'^\mu=\Lambda^\mu{}_{\nu}x^\nu$, the symmetric coordinates obey
\begin{equation}
q'^a=(\Lambda_q)^a{}_bq^b,
\qquad
\Lambda_q=U\Lambda U^\dagger,
\qquad
\Lambda_q^T\Lambda_q=I_4.
\label{eq:Lorentz-q}
\end{equation}
Preservation of the physical slice additionally requires
$\Lambda_q^*R_q=R_q\Lambda_q$.
The first condition expresses preservation of the complex bilinear form,
while the second ensures that transformed points continue to represent real
Minkowski events.

\section{Dual Variables and Fourier Representation}

\subsection{Wave covector and invariant phase}

Define $K_\mu=(-\omega/c,k_x,k_y,k_z)$ and
$\kappa_a=(U^{-1})^\mu{}_aK_\mu$. Then
\begin{equation}
\kappa_a=\frac12
\begin{bmatrix}
i\omega/c+k_x+k_y+k_z\\
i\omega/c-k_x+k_y-k_z\\
i\omega/c+k_x-k_y-k_z\\
i\omega/c-k_x-k_y+k_z
\end{bmatrix}_a.
\label{eq:kappa}
\end{equation}
The phase and invariant are
\begin{equation}
\kappa_aq^a=-\omega t+\bm k\cdot\bm x,
\qquad
\kappa_a\kappa^a=-\frac{\omega^2}{c^2}+|\bm k|^2.
\label{eq:spectral-invariant}
\end{equation}
The first equality confirms that the plane-wave phase is unchanged by the
coordinate transformation; the second shows that the light cone remains
$\kappa_a\kappa^a=0$.

\subsection{Four-momentum and de Broglie relation}

For $P_\mu=(-E/c,p_i)$, define
\begin{equation}
\pi_a=(U^{-1})^\mu{}_aP_\mu,
\qquad
\pi_aq^a=-Et+\bm p\cdot\bm x,
\qquad
\pi_a\pi^a=-m^2c^2.
\label{eq:momentum-q}
\end{equation}
The de Broglie relations are
\begin{equation}
\pi_a=\hbar\kappa_a,
\qquad
\kappa_a\kappa^a=-\left(\frac{mc}{\hbar}\right)^2.
\label{eq:deBroglie}
\end{equation}

\subsection{Fourier differentiation rule}

For $f(q)\propto\exp(i\kappa_aq^a)$,
\begin{equation}
\partial_a\longrightarrow i\kappa_a,
\qquad
\Box_q\equiv\delta^{ab}\partial_a\partial_b
\longrightarrow-\kappa_a\kappa^a.
\label{eq:Fourier-rule}
\end{equation}

\section{Dirac Field in Symmetric Spacetime}

\subsection{Transformed Dirac operator}

Let $\{\gamma^\mu,\gamma^\nu\}=2\eta^{\mu\nu}I_4$ and consider
$(\hbar\gamma^\mu\partial_\mu+mc)\psi=0$. Define
\begin{equation}
\Gamma^a=U^a{}_{\mu}\gamma^\mu
=\frac12\left(i\gamma^0+s_i^{\,a}\gamma^i\right).
\label{eq:Gamma}
\end{equation}
Then
\begin{equation}
(\hbar\Gamma^a\partial_a+mc)\Psi=0,
\qquad
\{\Gamma^a,\Gamma^b\}=2\delta^{ab}I_4,
\label{eq:Dirac-q}
\end{equation}
and $(\Gamma^a\partial_a)^2=\Box_qI_4$.
Consequently, multiplication by the conjugate massive first-order operator
recovers the Klein--Gordon equation for every spinor component.

In the Weyl representation,
\begin{equation}
\Gamma^a=\frac12
\begin{bmatrix}
0&iI_2+\bm s^{\,a}\cdot\bm\sigma\\
-iI_2+\bm s^{\,a}\cdot\bm\sigma&0
\end{bmatrix}.
\label{eq:Gamma-Pauli}
\end{equation}
Equation~\eqref{eq:Gamma-Pauli} makes the tetrahedral content explicit: each
symmetric direction selects one of four equiangular combinations of the
three Pauli matrices.

\subsection{Lagrangian and conserved current}

With $\bar\Psi=\Psi^\dagger(i\gamma^0)$, the free Lagrangian and current
are
\begin{align}
\mathcal L_D&=\frac{\hbar c}{2}
\left[\bar\Psi\Gamma^a\partial_a\Psi
-(\partial_a\bar\Psi)\Gamma^a\Psi\right]
+mc^2\bar\Psi\Psi,\label{eq:Dirac-L}\\
j_q^a&=-ic\,\bar\Psi\Gamma^a\Psi,
\qquad \partial_aj_q^a=0.\label{eq:Dirac-current}
\end{align}

\section{Gauge Coupling and QED in Symmetric Spacetime}

\subsection{Gauge potential and covariant derivative}

Define
\begin{equation}
\mathcal A_a=(U^{-1})^\mu{}_aA_\mu,
\qquad
\mathcal F_{ab}=\partial_a\mathcal A_b-\partial_b\mathcal A_a.
\label{eq:gauge-transform}
\end{equation}
For signed charge $e$, let
$D_a=\partial_a+(ie/\hbar)\mathcal A_a$. The QED Lagrangian is
\begin{equation}
\mathcal L_{\rm QED}=\frac{\hbar c}{2}
\left[\bar\Psi\Gamma^aD_a\Psi
-(D_a\bar\Psi)\Gamma^a\Psi\right]
+mc^2\bar\Psi\Psi
-\frac{1}{4\mu_0}\mathcal F_{ab}\mathcal F^{ab}.
\label{eq:QED-L}
\end{equation}
Here $D_a$ denotes the electromagnetic gauge-covariant derivative; it is not
the numerical differentiation matrix introduced later.

\subsection{Field equations and operator square}

It gives
\begin{equation}
(\hbar\Gamma^aD_a+mc)\Psi=0,
\qquad
\partial_a\mathcal F^{ab}=\mu_0\mathcal J_{\rm em}^{\,b},
\qquad
\partial_{[a}\mathcal F_{bc]}=0,
\label{eq:QED-equations}
\end{equation}
where
$\mathcal J_{\rm em}^{\,a}=-iec\,\bar\Psi\Gamma^a\Psi$.
With $\Sigma^{ab}=[\Gamma^a,\Gamma^b]/2$,
\begin{equation}
(\Gamma^aD_a)^2
=D_aD^a+\frac{ie}{2\hbar}\Sigma^{ab}\mathcal F_{ab}.
\label{eq:Pauli-coupling}
\end{equation}

\section{Maxwell Field and Tetrahedral Spin-1 Structure}

\subsection{Riemann--Silberstein evolution operator}

Define the Riemann--Silberstein vector $\bm F=\bm E+ic\bm B$, which is
distinct from the antisymmetric field tensor $\mathcal F_{ab}$. Maxwell's
equations become
\begin{equation}
\nabla\cdot\bm F=\frac{\rho}{\epsilon_0},
\qquad
\partial_{\parallel}\bm F+\nabla\times\bm F
=\frac{i}{\epsilon_0c}\bm J,
\end{equation}
where $\partial_{\parallel}=e_0^a\partial_a=-i\partial_t/c$ and
$\partial_i=e_i^a\partial_a=s_i^{\,a}\partial_a/2$.
For the cross-product matrix $[\bm v]_\times$, define
\begin{equation}
M^a=\frac12\left(I_3+[\bm s^{\,a}]_\times\right).
\label{eq:Maxwell-M}
\end{equation}
Then $M^a\partial_a=\partial_{\parallel}I_3+[\nabla]_\times$. With Cartesian spin-1
generators $(S_i)_{jk}=-i\epsilon_{ijk}$,
\begin{equation}
M^a=\frac12\left(I_3-is_i^{\,a}S_i\right).
\label{eq:M-spin1}
\end{equation}
The identity term advances the field along the common Hadamard time
direction, while the spin-1 term represents the spatial curl.

\subsection{Complete Maxwell operator}

To include Gauss's law and the evolution equations, define
\begin{equation}
\mathbb M^a=\frac12
\begin{bmatrix}
(\bm s^{\,a})^T\\
I_3+[\bm s^{\,a}]_\times
\end{bmatrix},
\qquad
\mathcal J_M=
\begin{bmatrix}
\rho/\epsilon_0\\
i\bm J/(\epsilon_0c)
\end{bmatrix}.
\label{eq:complete-M}
\end{equation}
The complete system is
\begin{equation}
\mathbb M^a\partial_a\bm F=\mathcal J_M.
\label{eq:Maxwell-complete}
\end{equation}
The first row of $\mathbb M^a$ supplies Gauss's law and its remaining three
rows supply the evolution equations; retaining both blocks is necessary for
the rectangular factorization below.

\subsection{Rectangular Clifford-Like Factorization}

Define
\begin{equation}
\mathbb N^a=\frac12
\begin{bmatrix}
\bm s^{\,a}&I_3-[\bm s^{\,a}]_\times
\end{bmatrix}.
\label{eq:Nmatrices}
\end{equation}
Then
\begin{equation}
\mathbb N^a\mathbb M^b+\mathbb N^b\mathbb M^a
=2\delta^{ab}I_3,
\label{eq:rectangular-Clifford}
\end{equation}
This anticommutator-like relation is rectangular because $\mathbb M^a$ maps
a three-component field to four Maxwell equations, whereas $\mathbb N^a$
maps the four-equation residual back to the field space.
and, because the derivatives commute,
\begin{equation}
(\mathbb N^a\partial_a)(\mathbb M^b\partial_b)=\Box_qI_3,
\qquad
(\Gamma^a\partial_a)^2=\Box_qI_4.
\label{eq:common-factorization}
\end{equation}

\subsection{Spectral Form and Unified Dirac--Maxwell Solver}

\subsubsection{Spectral inverses}

The free Dirac symbol is
$\mathcal D_D(\kappa)=i\hbar\Gamma^a\kappa_a+mc$, with inverse
\begin{equation}
\mathcal D_D^{-1}(\kappa)
=\frac{-i\hbar\Gamma^a\kappa_a+mc}
{\hbar^2\kappa_a\kappa^a+m^2c^2}.
\label{eq:Dirac-spectral}
\end{equation}
The complete Maxwell system gives
\begin{equation}
\widetilde{\bm F}
=-\frac{i\mathbb N^a\kappa_a}{\kappa_b\kappa^b}
\widetilde{\mathcal J}_M,
\qquad \kappa_b\kappa^b\ne0.
\label{eq:Maxwell-spectral}
\end{equation}
Both systems therefore have spin-dependent matrix numerators over scalar
relativistic propagation denominators.

\subsubsection{Discrete operators and preconditioning}

For compatible discrete derivatives $\mathsf D_a$, define
\begin{equation}
L_D=\hbar\Gamma^a\mathsf D_a+mcI_4,
\qquad
L_M=\mathbb M^a\mathsf D_a,
\qquad
L_M^\sharp=\mathbb N^a\mathsf D_a.
\end{equation}
The factorization motivates
\begin{equation}
L_M^\sharp L_M\simeq L_\Box I_3,
\qquad
P_M^{-1}\simeq L_\Box^{-1}L_M^\sharp.
\label{eq:preconditioner}
\end{equation}
Since $U^\dagger U=I_4$, the Hadamard transformation has two-norm
condition number one. Any computational advantage must arise from the
factorization, discretization, preconditioning, or reuse of common kernels.

\section{Weak-Field Einstein--Dirac--Maxwell System}
\label{sec:weak-edm}

This section treats gravity to first order in the metric perturbation while
retaining the electromagnetic coupling of the Dirac field. Quantities marked
by $(0)$ are evaluated on the flat symmetric background; quantities marked
by $(1)$ are their leading gravitational corrections. This ordering separates
the source of the linearized Einstein equation from the action of the
resulting weak metric on the matter fields.

The symmetric-spacetime formulation extends to weak gravitation by writing
\begin{equation}
g_{ab}(q)=\delta_{ab}+h_{ab}(q),
\qquad |h_{ab}|\ll1,
\qquad
g^{ab}=\delta^{ab}-h^{ab}+O(h^2).
\label{eq:weak-metric}
\end{equation}
Although the background metric has Euclidean form, $g_{ab}$ is a complex
bilinear metric on the physical slice. A real Lorentzian metric requires
\begin{equation}
h_{ab}^*=(R_q)^c{}_a(R_q)^d{}_b h_{cd}.
\label{eq:h-reality}
\end{equation}
Thus Eq.~\eqref{eq:weak-metric} is a representation of a conventional weak
Lorentzian metric, rather than a Euclidean gravitational theory.

\subsection{Linearized Einstein equation}

Define
\begin{equation}
h=\delta^{ab}h_{ab},
\qquad
\bar h_{ab}=h_{ab}-\frac12\delta_{ab}h.
\label{eq:trace-reverse}
\end{equation}
To first order, the connection and Ricci tensor are
\begin{align}
\Gamma^{a(1)}{}_{bc}
&=\frac12\delta^{ad}
\left(\partial_bh_{dc}+\partial_ch_{db}-\partial_dh_{bc}\right),
\label{eq:linear-connection}\\
R_{ab}^{(1)}
&=\frac12\left(
\partial_c\partial_a h^c{}_b
+\partial_c\partial_b h^c{}_a
-\Box_q h_{ab}-\partial_a\partial_bh
\right).
\label{eq:linear-Ricci}
\end{align}
Impose the harmonic gauge
\begin{equation}
\partial^a\bar h_{ab}=0.
\label{eq:harmonic-gauge}
\end{equation}
The linearized Einstein tensor then reduces to
$G_{ab}^{(1)}=-\Box_q\bar h_{ab}/2$, and Einstein's equation becomes
\begin{equation}
\Box_q\bar h_{ab}
=-\frac{16\pi G}{c^4}
\left(T_{ab}^{D(0)}+T_{ab}^{\mathrm{EM}(0)}
+T_{ab}^{\mathrm{ext}(0)}\right).
\label{eq:linear-Einstein-q}
\end{equation}
The superscript $(0)$ indicates that a strictly first-order gravitational
equation is sourced by stress tensors evaluated on the flat symmetric
background. Corrections to $T_{ab}$ proportional to $h_{ab}$ would contribute
to Einstein's equation only at the next gravitational order.
The harmonic constraint is compatible with Eq.~\eqref{eq:linear-Einstein-q}
provided the leading source obeys
\begin{equation}
\partial^aT_{ab}^{(0)}=0.
\label{eq:source-conservation}
\end{equation}
This conservation law follows from the coupled flat-background Dirac and
Maxwell equations when any external source is conserved as well.

The electromagnetic stress tensor on the flat symmetric background is
\begin{equation}
T_{ab}^{\mathrm{EM}(0)}
=\frac{1}{\mu_0}
\left(
\mathcal F_{ac}\mathcal F_b{}^c
-\frac14\delta_{ab}\mathcal F_{cd}\mathcal F^{cd}
\right).
\label{eq:TEM}
\end{equation}
The Dirac stress tensor is most unambiguously defined by metric variation,
\begin{equation}
T_{ab}^D=-\frac{2}{\sqrt{-g}}
\frac{\delta S_D}{\delta g^{ab}},
\label{eq:TD-variation}
\end{equation}
whose on-shell flat-background form, in the conventions of
Eq.~\eqref{eq:Dirac-L}, is
\begin{equation}
T_{ab}^{D(0)}=\frac{\hbar c}{2}
\left[
\bar\Psi\Gamma_{(a}D_{b)}\Psi
-(D_{(a}\bar\Psi)\Gamma_{b)}\Psi
\right].
\label{eq:TD-flat}
\end{equation}
For a quantized Dirac field in semiclassical gravity, $T_{ab}^{D(0)}$ in
Eq.~\eqref{eq:linear-Einstein-q} is replaced by the renormalized expectation
value $\langle\widehat T_{ab}^{D(0)}\rangle_{\rm ren}$
\cite{Kain2023}.

\subsection{Dirac field in the weak metric}

To distinguish background and curved matrices, set
$\Gamma_{(0)}^a\equiv\Gamma^a$, where $\Gamma^a$ was defined in
Eq.~\eqref{eq:Gamma}. In a symmetric tetrad gauge, the corresponding curved
matrices are
\begin{equation}
\Gamma_h^a
=\Gamma_{(0)}^a-\frac12h^a{}_b\Gamma_{(0)}^b+O(h^2),
\qquad
\{\Gamma_h^a,\Gamma_h^b\}=2g^{ab}I_4+O(h^2).
\label{eq:curved-Gamma}
\end{equation}
Let
\begin{equation}
\Omega_a^{(1)}
=\frac18\omega_{abc}^{(1)}[\Gamma_{(0)}^b,\Gamma_{(0)}^c],
\qquad
\omega_{abc}^{(1)}
=\frac12\left(\partial_ch_{ab}-\partial_bh_{ac}\right)
\label{eq:spin-connection}
\end{equation}
denote the first-order spin connection in this tetrad gauge. The curved,
electromagnetically coupled Dirac equation is
\begin{equation}
\left[
\hbar\Gamma_h^a
\left(D_a+\Omega_a^{(1)}\right)+mc
\right]\Psi=O(h^2).
\label{eq:weak-Dirac-full}
\end{equation}
Equivalently,
\begin{equation}
\left(\hbar\Gamma_{(0)}^aD_a+mc+\delta\mathcal D_h\right)\Psi=O(h^2),
\label{eq:weak-Dirac}
\end{equation}
where
\begin{equation}
\delta\mathcal D_h
=-\frac{\hbar}{2}h^a{}_b\Gamma_{(0)}^bD_a
+\hbar\Gamma_{(0)}^a\Omega_a^{(1)}.
\label{eq:Dirac-gravity-correction}
\end{equation}
The first term changes the local propagation frame, whereas the second is the
spin--gravity coupling. Correspondingly, the curved squared Dirac operator
contains both electromagnetic and curvature terms,
\begin{equation}
(\Gamma_h^a\mathcal D_a)^2
=\mathcal D_a\mathcal D^a
+\frac14R
+\frac{ie}{2\hbar}\Sigma^{ab}\mathcal F_{ab},
\label{eq:curved-Dirac-square}
\end{equation}
up to the adopted curvature-sign convention. Here
$\mathcal D_a=D_a+\Omega_a$ and
$\Sigma^{ab}=[\Gamma_h^a,\Gamma_h^b]/2$. Hence the flat scalar
factorization remains the principal part but is not an exact curved-space
identity.

\subsection{Maxwell field in the weak metric}

The generally covariant Maxwell equations are
\begin{equation}
\nabla_a\mathcal F^{ab}=\mu_0\mathcal J_{\rm em}^{\,b},
\qquad
\nabla_{[a}\mathcal F_{bc]}=0.
\label{eq:curved-Maxwell}
\end{equation}
In the curved Lorenz gauge $\nabla_a\mathcal A^a=0$, they imply
\begin{equation}
\nabla_c\nabla^c\mathcal A^a-R^a{}_b\mathcal A^b
=\mu_0\mathcal J_{\rm em}^{\,a},
\label{eq:curved-Maxwell-potential}
\end{equation}
with signs adjusted if the opposite curvature convention is used. Expanding
Eq.~\eqref{eq:curved-Maxwell} to first order gives
\begin{align}
\partial_a\mathcal F_{(0)}^{ab}
&-\partial_a\left(
h^a{}_c\mathcal F_{(0)}^{cb}+h^b{}_c\mathcal F_{(0)}^{ac}
\right)
+\frac12(\partial_ah)\mathcal F_{(0)}^{ab}
\nonumber\\
&=\mu_0\mathcal J_{\rm em}^{\,b}+O(h^2),
\label{eq:weak-Maxwell-expanded}
\end{align}
where
$\mathcal F_{(0)}^{ab}=\delta^{ac}\delta^{bd}\mathcal F_{cd}$ denotes the
field tensor with indices raised by the background metric. The subscript
$(0)$ is an order label and should not be confused with a temporal index.
Accordingly, the complete Riemann--Silberstein equation may be organized as
\begin{equation}
\mathbb M^a\partial_a\bm F
=\mathcal J_M+\mathcal S_h[\bm F]+O(h^2),
\label{eq:weak-Maxwell-RS}
\end{equation}
where $\mathcal S_h$ collects the first-order metric and connection terms.
Applying the complementary operator yields
\begin{equation}
\Box_q\bm F
=\mathbb N^a\partial_a
\left(\mathcal J_M+\mathcal S_h[\bm F]\right)+O(h^2).
\label{eq:weak-Maxwell-factor}
\end{equation}
Thus the rectangular factorization remains available as the flat principal
operator and as a natural preconditioner for the gravitationally perturbed
Maxwell system.

\subsection{Common spectral kernel and perturbative solution}

For compactness, define the total leading source
\begin{equation}
T_{ab}^{(0)}
=T_{ab}^{D(0)}+T_{ab}^{\mathrm{EM}(0)}+T_{ab}^{\mathrm{ext}(0)}.
\label{eq:total-leading-source}
\end{equation}
Fourier transformation of Eq.~\eqref{eq:linear-Einstein-q} gives
\begin{equation}
\widetilde{\bar h}_{ab}(\kappa)
=\frac{16\pi G}{c^4}
\frac{\widetilde T_{ab}^{(0)}(\kappa)}
{\kappa_c\kappa^c},
\qquad
\kappa^a\widetilde{\bar h}_{ab}=0,
\label{eq:gravity-spectral}
\end{equation}
subject to the usual causal prescription at $\kappa^2=0$. Equations
\eqref{eq:Dirac-spectral}, \eqref{eq:Maxwell-spectral}, and
\eqref{eq:gravity-spectral} show that gravity and electromagnetism share the
massless denominator $\kappa^2$, whereas the Dirac denominator is
$\hbar^2\kappa^2+m^2c^2$.

A consistent first-order gravitational calculation can be arranged as
\begin{equation}
\Psi^{(0)},\mathcal A^{(0)}
\longrightarrow
\left(\mathcal J_{\rm em}^{(0)},T_D^{(0)},T_{\rm EM}^{(0)}\right)
\longrightarrow
\bar h^{(1)}
\longrightarrow
\left(\Psi^{(1)},\mathcal A^{(1)}\right).
\label{eq:weak-coupling-chain}
\end{equation}
More explicitly,
\begin{align}
\left(\hbar\Gamma_{(0)}^aD_a+mc\right)\Psi^{(1)}
&=-\delta\mathcal D_h[h^{(1)}]\Psi^{(0)},
\label{eq:Dirac-first-correction}\\
\mathbb M^a\partial_a\bm F^{(1)}
&=\delta\mathcal J_M[\Psi^{(1)}]
+\mathcal S_h[h^{(1)},\bm F^{(0)}].
\label{eq:Maxwell-first-correction}
\end{align}
This ordering includes the action of the generated weak gravitational field
on the matter sectors without inserting second-order gravitational terms into
the linearized Einstein equation.

The combined principal operators are summarized by
\begin{equation}
\begin{aligned}
\Box_q\bar h_{ab}
&=-\frac{16\pi G}{c^4}T_{ab}^{(0)},\\
(\mathbb N^a\partial_a)(\mathbb M^b\partial_b)\bm F
&=\Box_q\bm F,\\
(\Gamma^a\partial_a)^2\Psi
&=\Box_q\Psi.
\end{aligned}
\label{eq:EDM-common-principal}
\end{equation}
Equation~\eqref{eq:EDM-common-principal} is a common flat-background
principal-operator structure, not an assertion that the complete curved
spin-$1/2$, spin-1, and spin-2 operators are identical. The lower-order
curvature couplings retain the distinct spin content of the three fields.

\section{Discussion}

The normalized Hadamard transformation provides a common tetrahedral
organization of spacetime coordinates, dual spectral variables, and
relativistic field operators. The same four sign vectors organize the
spin-$1/2$ matrices through $s_i^{\,a}\sigma_i$ and the spin-1 Maxwell
matrices through $s_i^{\,a}S_i$. The central algebraic result is the
rectangular identity
\begin{equation}
\mathbb N^a\mathbb M^b+\mathbb N^b\mathbb M^a
=2\delta^{ab}I_3,
\end{equation}
which applies to the complete Maxwell operator and places its factorization
in direct parallel with the Dirac Clifford algebra.

The weak-field extension adds the spin-2 gravitational sector without
altering Einstein gravity. In harmonic gauge, each component of the
trace-reversed metric perturbation has the same scalar principal operator
$\Box_q$ that results from the Dirac and Maxwell factorizations. This permits
the reuse of symmetric derivative kernels, Fourier variables, scalar wave
solvers, and block preconditioners. The equivalence is exact for the
flat-background principal parts. The complete weak-field equations remain
physically distinct because the Dirac equation contains the spin connection
and scalar-curvature term, while the Maxwell potential contains a Ricci
coupling and the gravitational field retains gauge constraints.

The formulation therefore suggests a computational architecture rather than
a new interaction or prediction. Existing Maxwell--Dirac algorithms provide
natural benchmarks \cite{Bao2004,Lorin2011,VandenBroeck2023}. A useful next
step is to compare conventional and symmetric-coordinate implementations of
the weak-field system using identical discretizations and boundary
conditions, measuring computational cost, iteration count, memory use, gauge
constraint preservation, and sensitivity to the curvature corrections.

\section{Conclusions}

A symmetric representation of Minkowski spacetime has been developed using a
normalized fourth-order Hadamard transformation. The resulting complex
coordinates distribute time and space symmetrically among four components
whose spatial sign vectors form a regular tetrahedron. These directions
organize the transformed Dirac and Riemann--Silberstein Maxwell operators
through their spin-$1/2$ and spin-1 representations. The complete Maxwell
operator satisfies a rectangular Clifford-like identity that, together with
the Dirac Clifford algebra, produces parallel first-order factorizations of
the scalar relativistic wave operator.

The formulation has been extended consistently to the weak-field
Einstein--Dirac--Maxwell system. In harmonic gauge, the linearized Einstein
equation is governed by the same symmetric wave operator $\Box_q$, and its
source is the leading Dirac and electromagnetic stress tensor. The weak
gravitational field acts back on the Dirac and Maxwell sectors through the
perturbed tetrad, spin connection, metric, and curvature. Consequently, the
three field sectors share a common flat-background principal and spectral
kernel, while their curvature corrections preserve their different spin and
gauge structures. The principal significance of this extension therefore lies in providing a unified representation and a potentially common geometric, algebraic, and computational framework.

\end{document}